# A SPectroscopic Survey of Biased Halos in the Reionization Era (ASPIRE): JWST Discovers an Overdensity around a Metal Absorption-selected Galaxy at $z \sim 5.5$

Yunjing Wu[1,2], Feige Wang[2], Zheng Cai[1], Xiaohui Fan[2], Kristian Finlator[3,4], Jinyi Yang[2,19], Joseph F. Hennawi[5,6], Fengwu Sun[2], Jaclyn B. Champagne[2], Xiaojing Lin[1], Zihao Li[1], Zuyi Chen[2], Eduardo Bañados[7,8], George D. Becker[9], Sarah E. I. Bosman[7,10], Gustavo Bruzual[11], Stephane Charlot[12], Hsiao-Wen Chen[13], Jacopo Chevallard[14], Anna-Christina Eilers[15,20], Emanuele Paolo Farina[16], Xiangyu Jin[2], Hyunsung D. Jun[17], Koki Kakiichi[5,18], Mingyu Li[1], Weizhe Liu[2], Maria A. Pudoka[2], Wei Leong Tee[2], Zhang-Liang Xie[7], and Siwei Zou[1]

[1] Department of Astronomy, Tsinghua University, Beijing 100084, People's Republic of China; yj-wu19@mails.tsinghua.edu.cn
[2] Steward Observatory, University of Arizona, 933 N Cherry Avenue, Tucson, AZ 85721, USA
[3] New Mexico State University, Las Cruces, 88003 NM, USA
[4] Cosmic Dawn Center (DAWN), Niels Bohr Institute, University of Copenhagen / DTU-Space, Technical University of Denmark, Denmark
[5] Department of Physics, Broida Hall, University of California, Santa Barbara, CA 93106-9530, USA
[6] Leiden Observatory, Leiden University, P.O. Box 9513, NL-2300 RA Leiden, The Netherlands
[7] Max Planck Institut für Astronomie, Königstuhl 17, D-69117, Heidelberg, Germany
[8] The Observatories of the Carnegie Institution for Science, 813 Santa Barbara Street, Pasadena, CA 91101, USA
[9] Department of Physics & Astronomy, University of California, Riverside, CA 92521, USA
[10] Institut für Theoretische Physik, Universität Heidelberg, Philosophenweg 16, D-69120 Heidelberg, Germany
[11] Institute of Radio Astronomy and Astrophysics, National Autonomous University of Mexico, San José de la Huerta 58089 Morelia, Michoacán, México
[12] Sorbonne Université, CNRS, UMR7095, Institut d'Astrophysique de Paris, F-75014, Paris, France
[13] Department of Astronomy & Astrophysics, The University of Chicago, 5640 S. Ellis Avenue, Chicago, IL 60637, USA
[14] Department of Physics, University of Oxford, Denys Wilkinson Building, Keble Road, Oxford OX1 3RH, UK
[15] MIT Kavli Institute for Astrophysics and Space Research, 77 Massachusetts Avenue, Cambridge, MA 02139, USA
[16] Gemini Observatory, NSF's NOIRLab, 670 N A'ohoku Place, Hilo, HI 96720, USA
[17] SNU Astronomy Research Center, Seoul National University, 1 Gwanak-ro, Gwanak-gu, Seoul 08826, Republic of Korea
[18] Cosmic Dawn Center (DAWN), Niels Bohr Institute, University of Copenhagen, Jagtvej 128, København N, DK-2200, Denmark

Received 2023 July 14; revised 2023 September 5; accepted 2023 September 23; published 2023 October 19

## Abstract

The launch of JWST opens a new window for studying the connection between metal-line absorbers and galaxies at the end of the Epoch of Reionization. Previous studies have detected absorber–galaxy pairs in limited quantities through ground-based observations. To enhance our understanding of the relationship between absorbers and their host galaxies at $z > 5$, we utilized the NIRCam wide-field slitless spectroscopy to search for absorber-associated galaxies by detecting their rest-frame optical emission lines (e.g., [O III] + H$\beta$). We report the discovery of a Mg II-associated galaxy at $z = 5.428$ using data from the JWST ASPIRE program. The Mg II absorber is detected on the spectrum of quasar J0305–3150 with a rest-frame equivalent width of 0.74 Å. The associated galaxy has an [O III] luminosity of $10^{42.5}$ erg s$^{-1}$ with an impact parameter of 24.9 pkpc. The joint Hubble Space Telescope–JWST spectral energy distribution (SED) implies a stellar mass and star formation rate of $M_* \approx 10^{8.8}\ M_\odot$, star-formation rate $\approx 10\ M_\odot$ yr$^{-1}$. Its [O III] equivalent width and stellar mass are typical of [O III] emitters at this redshift. Furthermore, connecting the outflow starting time to the SED-derived stellar age, the outflow velocity of this galaxy is $\sim 300$ km s$^{-1}$, consistent with theoretical expectations. We identified six additional [O III] emitters with impact parameters of up to $\sim 300$ pkpc at similar redshifts ($|dv| < 1000$ km s$^{-1}$). The observed number is consistent with that in cosmological simulations. This pilot study suggests that systematically investigating the absorber–galaxy connection within the ASPIRE program will provide insights into the metal-enrichment history in the early Universe.

*Unified Astronomy Thesaurus concepts:* Quasar absorption line spectroscopy (1317); Circumgalactic medium (1879); High-redshift galaxies (734)

## 1. Introduction

The circumgalactic medium (CGM), diffuse gas surrounding galaxies and inside their virial radii, plays an important role in our understanding of galaxy formation and evolution (Tumlinson et al. 2017). However, CGM is nearly invisible and difficult to be detected directly. Intervening metal-line absorbers, detected along the lines of sight of quasars, is a unique tracer for CGM gas in the early Universe. For example, the MUSE Analysis of Gas around Galaxies (MAGG) team has identified 220 C IV absorbers from 28 quasar sight lines at $3 < z < 4.5$ (Galbiati et al. 2023). Similarly, at higher redshifts ($z \gtrsim 5$), Chen et al. (2017), Becker et al. (2019), Cooper et al. (2019), Zou et al. (2021), and D'Odorico et al. (2022) have conducted statistically meaningful surveys of O I, Mg II, C IV, and Si IV absorbers. Through comprehensive searches, the XQR-30 group has also provided a catalog of 778 absorbers at $2 < z < 6.5$ (Davies et al. 2023). Detecting metal absorbers will help us to understand the details of the chemical enrichment of gaseous reservoirs surrounding galaxies.

---

[19] Strittmatter Fellow.
[20] Pappalardo Fellow.

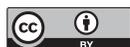

Theoretical models show that early star formation and associated feedback processes are responsible for enriching the early Universe with metals (Greif et al. 2010; Wise et al. 2012; Sorini et al. 2020). The absorber–galaxy correlation, directly linking galaxies and metal-enriched gas, is thus necessary to study the physical conditions of galaxies dominating the metal enrichment in the early Universe. Some cosmological simulations suggest that the typical stellar mass ($M_*$) of absorber host galaxies ranges from $\sim 10^7$ to $10^{9.5}$ $M_\odot$, with impact parameters (projected distance) ranging from a few to $\sim 150$ pkpc (Keating et al. 2016). At higher redshift, more recent simulations, such as the Technicolor Dawn simulations, suggest that less massive galaxies (with $M_* \lesssim 10^8 M_\odot$) could be responsible for metal absorbers at $z > 5$, with impact parameters $\leqslant 300$ pkpc (Finlator et al. 2020). Measurements of the impact parameters and star-formation rates (SFRs), stellar masses ($M_*$), and halo mass ($M_h$) are proposed to be useful for constraining the detailed physical process of metal enrichment (e.g., Oppenheimer et al. 2009; Finlator et al. 2013; Hirschmann et al. 2013). Therefore, identifying and characterizing galaxies hosting the metal-absorbing gas at high redshifts and studying their properties is crucial for understanding the early chemical enrichment process.

Observationally, identifying the host galaxies of metal absorbers at high redshifts has been proven to be challenging with ground-based observations. To date, only $\sim 10$ absorber–galaxy pairs have been identified at $z > 5$ (Cai et al. 2017; Díaz et al. 2021; Kashino et al. 2023b). Díaz et al. (2014, 2015) investigated the projected distribution of galaxies around two high column density C IV absorbers at $z \sim 5.7$ by searching for Ly$\alpha$ emitters (LAEs) around them. The closest galaxy to the C IV absorbers lies at $\sim 200$ pkpc from the C IV absorber along the line of sight of the $z = 6.3$ quasar J1030 + 0524. This galaxy has an SFR$_{UV} \approx 10$ $M_\odot$ yr$^{-1}$ and is at the high-mass end of the LAEs at these redshifts. These studies suggest that the high-C IV column density and C IV absorbers with high column density are either produced by large-scale outflows from relatively massive galaxies or outflows from undetected dwarf galaxies at closer distances. To explore galaxies at the luminosity faint end, Díaz et al. (2021) conducted deep Very Large Telescope (VLT)-MUSE observations of this sight line and identified a 4 times fainter LAE at a projected distance of $\sim 11$ pkpc to the C IV absorber. Cai et al. (2017) used deep Hubble Space Telescope (HST) narrowband observations to probe C IV-associated LAEs candidates at $z \sim 5.5$. They identified a candidate LAE with SFR$_{UV} \simeq 4$ $M_\odot$ yr$^{-1}$ with a projected impact parameter of 42 pkpc from the absorber. However, they cannot rule out that this LAE candidate could be an O II emitter at a much lower redshift because of the lack of spectroscopic observations. ALMA can also be used to search for UV-faint galaxies around metal absorbers and is not affected by dust obscuration. Wu et al. (2021) found one [C II] emitter candidate could be responsible for an O I absorber at $z \sim 6$. For highly ionized absorbers, Kashino et al. (2023b) reported two ALMA-detected [C II] emitters are associated with a C IV absorber at $z \approx 5.7$.

NIRCam/wide-field slitless spectroscopy (WFSS) on JWST has been proven to be highly effective in enabling galaxy surveys at $z \gtrsim 5.5$ by detecting strong rest-frame optical emission lines (e.g., [O III], H$\beta$; e.g., Sun et al. 2022, 2023; Kashino et al. 2023a; Oesch et al. 2023). Because of the large survey area and high sensitivity provided by WFSS observations, it is now possible to crossmatch galaxies and absorbers in the early Universe. For example, the Emission-line Galaxies and Intergalactic Gas in the Epoch of Reionization (EIGER) project presented early JWST observations linking [O III], H$\alpha$, and He I emitters with metal absorbers at $z \gtrsim 5.5$ (Kashino et al. 2023a; Bordoloi et al. 2023).

In this Letter, we report the discovery of an absorber–galaxy pair at $z = 5.428$ along the line of sight of quasar J0305–3150 identified with the A SPectroscopic Survey of Biased Halos In the Reionization Era (ASPIRE) program (Wang et al. 2023; Yang et al. 2023). Remarkably, the absorber host galaxy also traces galaxy overdensity. We named the host galaxy as ASPIRE-J0305M31-O3-038 (O3-038) in this Letter.

The Letter is structured as follows. In Section 2, we summarize the data used in this Letter and present the details of data reduction. We present the absorber searching, host galaxy identification, and the spatial distribution of the galaxies associated with the absorber host galaxy in Section 3. We compare our measurements with both the observational and theoretical studies in Section 4. Finally, we summarize our findings in Section 5. Throughout this Letter, we assume a flat $\Lambda$CDM cosmological model with $\Omega_M = 0.3$, $\Omega_\Lambda = 0.7$, and $H_0 = 70$ km s$^{-1}$ Mpc$^{-1}$. In this cosmological model, an angular size of 1″ corresponds to a physical scale of 6.024 pkpc at $z = 5.4$.

## 2. Target Selection, Observations, and Data Reduction

The quasar, J0305–3150 at $z = 6.61$, was discovered by Venemans et al. (2013) and covered by extensive multi-wavelength observations. To investigate the environment around this quasar, Farina et al. (2017) conducted deep MUSE observations to detect LAEs. Champagne et al. (2023) used deep HST imaging to find Lyman break galaxies surrounding this quasar. In this work, we select J0305–3150 as our pilot field to demonstrate the application of JWST observations in conjunction with existing data for investigating early metal enrichment.

### 2.1. JWST Observations

The JWST/NIRCam (Rieke et al. 2023) WFSS were obtained by Program #2078 (PI: F. Wang) with grism spectroscopic observations in F356W and imaging observations in F115W, F200W, and F356W. The on-source grism exposure time is 2834 s. The direct-imaging exposure times are 472, 2800, and 472 s in the F115W, F200W, and F356W filters, respectively. The data were reduced using the combination of the standard `JWST` pipeline (Bushouse et al. 2022)[21] v1.8.3 and some custom scripts as detailed in Wang et al. (2023) and Yang et al. (2023). We use calibration reference file version of "jwst_1015.pmap". We refer readers to Wang et al. (2023) and Yang et al. (2023) for a more detailed description of the process.

### 2.2. X-SHOOTER Observations

The VLT/X-SHOOTER NIR spectroscopy was obtained through Program ID: 098.B-0537(A) (PI: Farina) with a resolving power of $R \sim 8100$ and an exposure time of 16,

---
[21] https://github.com/spacetelescope/jwst

800 s. The data are reduced with `PypeIt`[22] (Prochaska et al. 2020) and presented in Schindler et al. (2020). We briefly recap the reduction procedures below. Sky subtraction was performed based on a standard ABBA method. We obtained the wavelength solutions by fitting the observed sky OH lines. The 1D spectrum was optimally extracted following the optimal-extraction method (Horne 1986). After that, the flux calibrations and telluric-absorption corrections were performed. For more detailed information, we refer readers to Schindler et al. (2020). We normalized the quasar continuum using `Linetools`[23] by manually adding knots for a spline fitting.

### 2.3. HST Observations

HST imaging observations of the field around quasar J0305 were obtained from the program GO #15064 (PI: Casey) for investigating the environment surrounding quasars at $z \approx 6$ (Champagne et al. 2023). Five bands were observed: F606W, F814W, F105W, F125W, and F160W. The exposure times are 2115.0, 2029.7, 2611.8, 2611.8, and 2611.8 s, respectively. All HST data were reduced using the standard `drizzlepac`[24] pipeline and with the astrometry registered to the JWST images. All the HST and JWST images are point-spread function (PSF)-matched to the F160W filter. Photometry is performed with `SourceXtractor++`[25] (Bertin et al. 2020; Kümmel et al. 2020) on the PSF-matched images. More details of the reduction procedures and the photometry measurements are presented in Champagne et al. (2023).

## 3. Analysis and Results

### 3.1. Mg II-absorber Searching and Voigt-profile Fitting

We briefly summarize our Mg II-absorber searching method in what follows, and more details on the method will be presented in a forthcoming paper (Y. Wu et al. 2023, in preparation.). Following Matejek & Simcoe (2012) and Chen et al. (2017), we first generated a normalized kernel using two Gaussian functions with the separation of the intrinsic Mg II-doublet interval (~751 km s$^{-1}$). The FWHM of these two Gaussian profiles ranges from 37.5 (the minimal resolution element) to 150 km s$^{-1}$, corresponding to reasonable line widths (Chen et al. 2017). The X-Shooter spectrum was convolved with the matched filter. We identify Mg II absorbers from the convolved spectrum by identifying peaks with a signal-to-noise Ratio (S/N) threshold of 2.5. We then visually inspected all identified Mg II absorber candidates to remove artifacts caused by sky-line residuals. Two Mg II absorbers passed the visual inspection and are identified along this sight line. This Mg II absorber has a typical rest-frame equivalent width (REW) of Mg II absorbers at $z = 5$–6 (Chen et al. 2017).

To obtain the detailed physical parameters from observed Mg II-absorption systems, we used `VoigtFit` to fit the absorption lines[26] (Krogager 2018). We need he initial guesses of the column densities ($\log(N_{\rm Mg\,II}/\rm cm^{-2}) = 13$) and Doppler parameters ($b = 10$ km s$^{-1}$) to make a fitting. The absorption-line fitting results for our target are shown in the bottom left panel of Figure 1. The derived $\log N$, $b$, and the REWs are listed in Table 1. We also use the "apparent optical depth method" (AODM; Savage & Sembach 1991) developed in `linetools` to check the fitting results. The column density measured using AODM is $\log(N_{\rm Mg\,II}/\rm cm^{-2}) = 13.43 \pm 0.05$, which is consistent with our Voigt fitting results.

### 3.2. The Host Galaxy and Environment of a Mg II-absorber

Metal-line absorbers are thought to be from metal-polluted gas blown out by their host galaxies. In this scenario, the velocity offset between the absorbers and their host galaxies is typically within $\Delta v \lesssim \pm 200$ km s$^{-1}$ (e.g., Steidel et al. 2010; Keating et al. 2016; Díaz et al. 2021). The velocity offsets between Mg II-absorbers and [O III] emitters are calculated as $\Delta v([\rm O\,III] - Mg\,II) = (z_{\rm [O\,III]} - z_{\rm Mg\,II})/(1 + z_{\rm Mg\,II}) \times c$, where $z_{\rm Mg\,II}$ and $z_{\rm [O\,III]}$ denote the redshifts of the Mg II absorber and [O III] emitters, respectively, and $c$ is the speed of light. To identify the host galaxies of the Mg II absorbers, we matched redshifts of the Mg II absorbers with those of [O III] emitters reported in Wang et al. (2023). Along the line of sight of J0305–3150, we identified an [O III] emitter, ASPIRE-J0305M31-O3-038 (hereafter O3-038), located at the exact redshift of the Mg II absorber at $z = 5.428 \pm 0.003$ ($dv = -2.9 \pm 140$ km s$^{-1}$). The JWST spectrum of the [O III] emitter and the absorption spectrum of the Mg II absorber are shown in the left panel of Figure 1. As shown in the right panel of Figure 1, the observed impact parameter between the Mg II absorber and the [O III] emitter is 4″.1, corresponding to 24.9 pkpc at $z = 5.428$.

We compare the observed impact parameter with the expected halo virial radius to determine if this Mg II absorber belongs to CGM gas of O3-038. To obtain the virial radius and physical properties of the galaxy, we performed SED fitting using photometry from HST and JWST with the Bayesian Analysis of Galaxy SEDs codes, `BEAGLE`[27] (Chevallard & Charlot 2016), following the procedures described in Lin et al. (2023), Whitler et al. (2023), and Chen et al. (2023). We note that, to get precise SED estimation, it is necessary to subtract the emission line contributions from the photometry. We measured the rest-frame [O III] equivalent widths from the grism observations. The measured [O III] flux is $f_{\rm [O\,III]} = (1.013 \pm 0.084) \times 10^{-17}$ erg s$^{-1}$ cm$^{-2}$. We then estimated the continuum level by subtracting the line flux from the broad-band photometry and obtained EW$_{\rm [O\,III]} = 490 \pm 80$ Å. From the SED fitting, we derive the stellar mass, SFR, and stellar age of this galaxy to be $M_* \simeq 10^{8.8^{+0.2}_{-0.4}}\,M_\odot$, SFR $\simeq 9.7^{+4.3}_{-2.7}\,M_\odot$ yr$^{-1}$, and age $= 79^{+79}_{-57}$ Myr, respectively (also see Appendix). We note that we do not have H$\beta$ detection for O3-038 due to a low transmission at the expected wavelength (Figure 2 in Wang et al. 2023). Thus, we do not have the comparison between H$\beta$-derived and SED-derived SFRs. Further, by utilizing the stellar mass to halo mass relation employed in Ma et al. (2018), we estimate the halo mass of $\sim 1.2 \times 10^{11}\,M_\odot$. At $z \sim 5.5$, the corresponding virial radius is $\sim 25$ pkpc. We also obtain a consistent estimation using the empirical relation reported in Sun & Furlanetto (2016). We note that the estimated virial radius is consistent with the measured impact parameter. This simple estimation suggests that the O3-038 is the host galaxy of metal absorber at $z = 5.428$.

---

[22] https://pypeit.readthedocs.io/en/release/
[23] https://linetools.readthedocs.io/en/latest/index.html
[24] https://github.com/spacetelescope/drizzlepac
[25] https://github.com/astrorama/SourceXtractorPlusPlus
[26] https://voigtfit.readthedocs.io/en/latest/

[27] https://www.iap.fr/beagle/

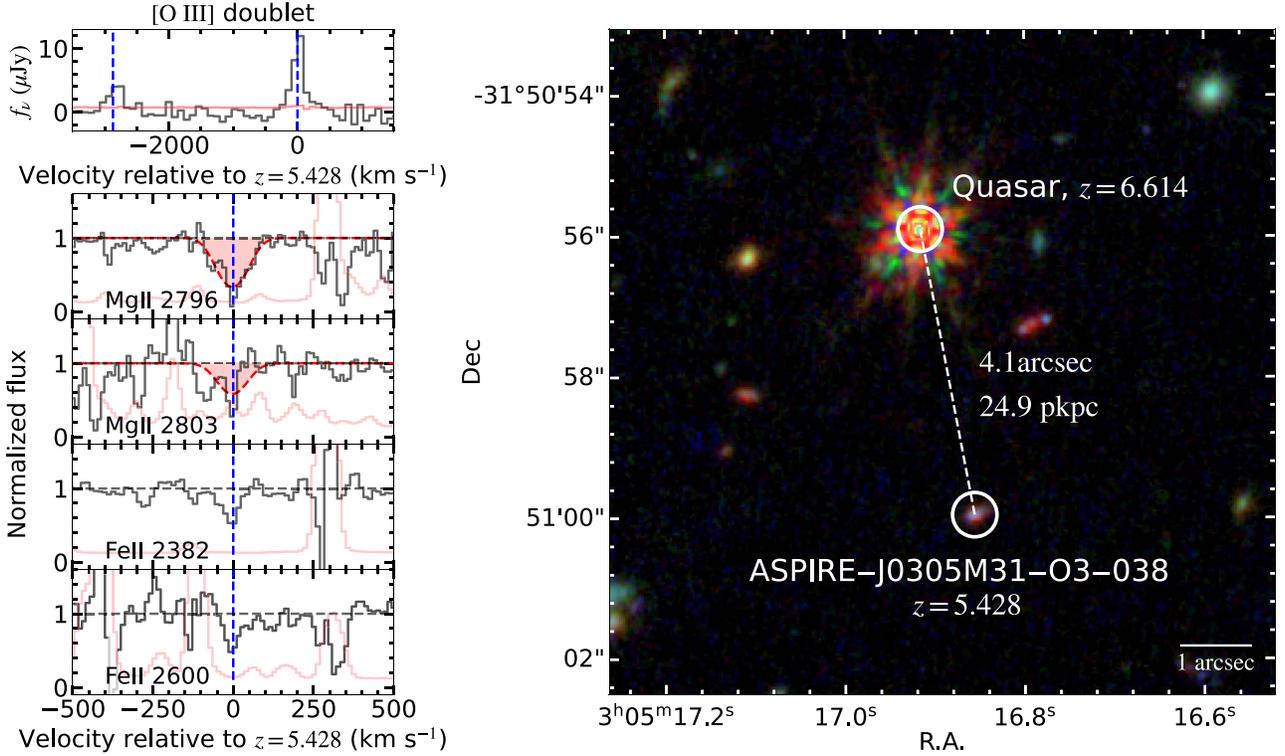

**Figure 1.** Top left: the JWST/NIRCam WFSS spectrum of ASPIRE-J0305M31-O3-038. The error spectrum is shown in pink lines. Bottom left: normalized X-SHOOTER spectrum of the quasar J0305 with Mg II $\lambda\lambda$2796, 2803 and Fe II $\lambda$ 2382, 2600 absorptions at $z = 5.4284$. Red shaded regions denote the best-fit Voigt profile. Dashed blue lines indicate the redshift. Right: JWST/NIRCam composite the red, green, and blue map of the quasar field with the pixel scale of 0″03 (blue: F115W, green: F200W, red: F356W). White circles denote the location of ASPIRE-J0305M31-O3-038 and the quasar J0305. We note that the Mg II absorber is in front of the quasar with the redshift of $z = 5.428$. The impact parameter between the Mg II absorber and ASPIRE-J0305M31-O3-038 is 4″1, corresponding to 24.9 pkpc at $z = 5.428$.

**Table 1**
Physical Properties of ASPIRE-J0305M31-O3-038

| Galaxy | ASPIRE-J0305M31-O3-038 |
|---|---|
| R.A. (deg) | 46.32023 |
| Decl. (deg) | −31.84999 |
| $z_{\rm spec}$ | 5.428 ± 0.003 |
| impact parameter (pkpc) | 25 |
| Absorption Properties | |
| $z_{\rm abs}$ | 5.4284 ± 0.0001 |
| $\log(N_{\rm Mg\,II}/{\rm cm}^{-2})$ | 13.41 ± 0.06 |
| $b$ (km s$^{-1}$) | 56.5 ± 9.3 |
| REW(Mg II$\lambda$2796) (Å) | 0.74 ± 0.10 |
| REW(Mg II$\lambda$2803) (Å) | 0.53 ± 0.19 |
| Photometric Properties | |
| F356W (AB mag) | 25.71 ± 0.04 |
| Physical parameters | |
| $f$([O III])[a] (10$^{-17}$ erg s$^{-1}$ cm$^{-2}$) | 1.013 ± 0.084 |
| $L$([O III])[a] (10$^{42}$ erg s$^{-1}$) | 3.30 ± 0.51 |
| EW([O III])[a] (Å) | 490 ± 80 |
| SFR[b] ($M_\odot$ yr$^{-1}$) | $9.7^{+4.3}_{-2.7}$ |
| $\log[M_{\rm star}/(M_\odot)]$[b] | $8.8^{+0.2}_{-0.4}$ |
| $t_{\rm age}$[b] (Myr) | $79^{+79}_{-57}$ |

**Notes.**
[a] For [O III] doublet $\lambda\lambda$4959, 5007.
[b] Stellar mass and age are derived based on a constant star formation history in BEAGLE SED fitting (see Appendix).

We further estimate the one-dimensional velocity dispersion ($\sigma_{\rm 1d}$) and the mass scale ($M_{\rm h\_overdensity}$) for this overdensity, following the procedure described in Ferragamo et al. (2020) and Evrard et al. (2008). We estimate $\sigma_{\rm 1d}$ by calculating the variance of the line-of-sight velocities (Evrard et al. 2008), i.e., $\sigma_{\rm 1d}^2 = \frac{1}{N_p}\sum_{i=1}^{N_p}(v_i - \bar{v})^2$, where $v_i$ is the velocity of halo member $i$, $N_p$ is the number of halo members, and $\bar{v}$ is the mean velocity of these member galaxies. The calculated $\sigma_{\rm 1d} = 346.4$ km s$^{-1}$. For the mass scale, we used the scaling relations between $\sigma_{\rm 1d}$ and the halo mass (Ferragamo et al. 2020; Munari et al. 2013), i.e., $\frac{\sigma_{\rm 1d}}{\rm km\,s^{-1}} = A\left[\frac{h(z)M_h}{10^{15}M_\odot}\right]^\alpha$, where $A = 1177.0$ km s$^{-1}$, and $\alpha = 1/3$ (Munari et al. 2013). Therefore, the estimated halo mass of this overdensity is $M_{\rm h\_overdensity} = 4.1 \times 10^{12} M_\odot$.

Interestingly, we found that O3-038 resides in an overdense environment. In the field of quasar J0305, six additional [O III] emitters were detected with WFSS within a velocity offset of ±1000 km s$^{-1}$ to $z = 5.428$. Our detection limit is $L_{\rm [O\,III]} > 10^{42.3}$ erg s$^{-1}$. To determine the significance of this overdensity, we compared our observations with the expected number of [O III] emitters in a random field. We calculated the mean number density using the [O III] luminosity function reported by Sun et al. (2023) and Matthee et al. (2023). The mean number density is $n(L_{\rm [O\,III]} > 10^{42.3}$ erg s$^{-1}) \approx 1.873 \times 10^{-3}$ cMpc$^{-3}$. In a comparable survey volume to our case (a cube with an area of $\approx$5.5 arcmin$^2$ and height of ±1000 km s$^{-1}$, corresponding to a survey volume of $\approx$607 cMpc$^3$), the number of randomly detected [O III] emitters brighter than our detection limit is $\sim$1.1. Therefore, we observe a number density 6 times higher than expected in a blank field. This suggests that the Mg II absorber is associated with an overdense environment. The spatial and redshift distributions of these [O III] emitters are shown in Figure 2.

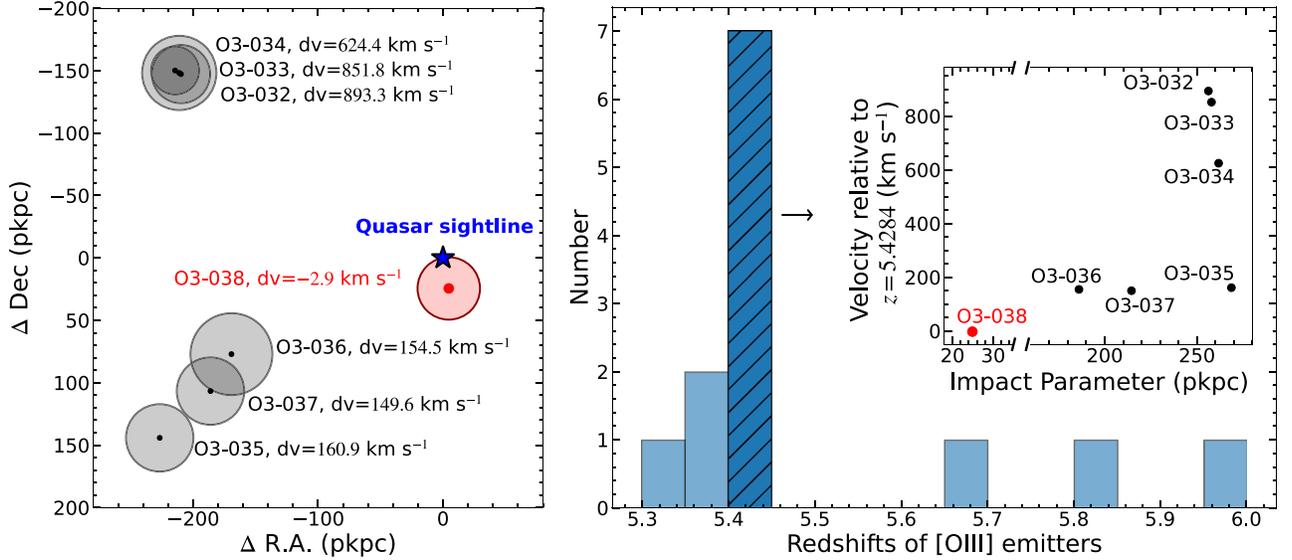

**Figure 2.** Left: spatial distribution of [O III] emitters relative to the quasar. The location of the quasar is shown with a blue star. Dots represent the locations of [O III] emitters, while the shaded circles correspond to the virial radii of galaxies estimated from their best SED-derived stellar masses. Right: the redshift distribution of [O III] emitters at $z < 6$ in this quasar field (J0305). The hatched bar indicates [O III] emitters clustering at $z \simeq 5.4$. Targets at other redshifts ($|dv| > 1000$ km s$^{-1}$) are shown in blank bars, $\Delta v({\rm [O\,III]} - {\rm Mg\,II}) = (z_{\rm [O\,III]} - z_{\rm Mg\,II})/(1 + z_{\rm Mg\,II}) \times c$. The velocity offsets and impact parameters are shown in the top right panel. The Mg II–associated galaxy is highlighted in red.

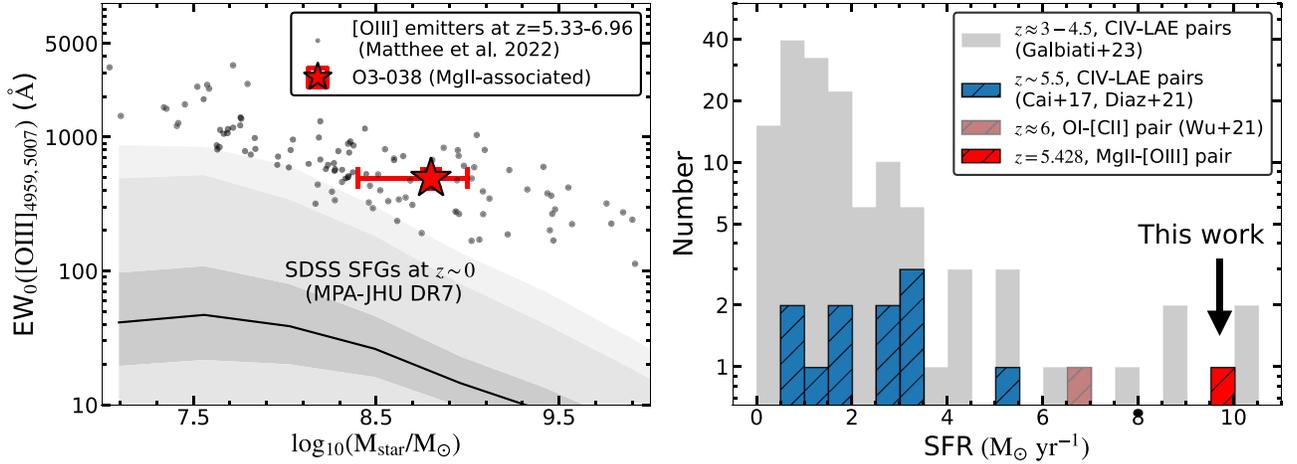

**Figure 3.** Left: [O III] EWs and derived stellar masses at $z \sim 6$, compared to those of local galaxies (shaded region; MPA-JHU catalog (Kauffmann et al. 2003; Brinchmann et al. 2004)). The measurement of the absorber-associated [O III] emitter is shown as the red star. Gray dots indicate results obtained from the EIGER sample (Matthee et al. 2023). Right: SFR distribution of metal-absorber-selected galaxies. Galaxies at $z > 5$ are marked as hatched bars. At $z < 5$, the C IV-associated LAEs selected from the MAGG survey (Galbiati et al. 2023) are shown in gray. The blue bars show the C IV-associated LAEs at $z > 5$ (Cai et al. 2017; Díaz et al. 2021). At $z \approx 6$, one ALMA-detected O I-associated emitter is shown in pink (Wu et al. 2021). The [O III] emitter detected with JWST is shown in red.

## 4. Discussion

### 4.1. Properties of the Absorption-selected Galaxy

The primary goal of this work is to investigate the nature of galaxies that host metal line absorbers. The left panel of Figure 3 shows the measurements of [O III] EWs and $M_*$ from galaxies in both high redshift (Matthee et al. 2023) and local Universe (Kauffmann et al. 2003; Brinchmann et al. 2004). Generally, lower-mass galaxies exhibit higher [O III] EW, which will be expected if they have relatively higher specific SFRs (Mannucci et al. 2010; Tang et al. 2019; Endsley et al. 2021). We find that the [O III] EW and stellar mass of O3-038 are consistent with the general population of [O III] emitters at $z \sim 6$. Similar to other [O III] emitters, O3-038 exhibits a larger EW than local star-forming galaxies (SFGs) at matched stellar mass. Similar to other [O III] emitters, O3-038 exhibits a larger EW than local SFGs at matched stellar mass. The observed large [O III] equivalent width (EW$_{\rm [O\,III]} = 490 \pm 80$) suggests the presence of the stellar population with the age of $\sim 50$ Gyr (Tang et al. 2019). Such young stellar populations could generate hard ionizing photons. Consequently, the observed [O III] EW will be relatively higher than local SFGs, given a specific stellar mass.

We also examine the similarity between this Mg II-selected [O III] emitter and other metal-absorber-selected galaxies.

Because different galaxies were identified by different tracers, e.g., Lyα or [C II] emission, we compare their SFRs for consistency. For example, for metal-related LAEs, we convert their measured Lyα luminosity to the SFRs according to the relation in Ouchi et al. (2020). For metal-selected [C II] emitters, we followed the $L_{\rm [C\,II]}$–SFR relation reported by Schaerer et al. (2020). Figure 3 right panel shows our results. We find that ASPIRE-J0305M31-O3-038 has a higher SFR than the majority of the absorber-selected galaxy sample. We note that there will be internal systematic offsets between different SFR tracers.

### 4.2. Absorber–Galaxy Cross-correlation Function

Determining the correlation function between galaxies and metal-line absorbers is critical for our understanding of which galaxies are responsible for the metal enrichment in the early Universe (Keating et al. 2016; Meyer et al. 2019; Finlator et al. 2020). We calculate the galaxy number excess relative to the blank field around absorbers as a function of impact parameters ($r$):

$$\xi_{\rm gal-abs}(r) = \frac{1}{n_0} \frac{N(r)}{\Delta V(r)} - 1, \quad (1)$$

where $n_0$ is the comoving number density of [O III] emitters with $L_{\rm [O\,III]}$ brighter than our detection limit, $N(r)$ indicates the number of [O III] emitters around an absorber, and $\Delta V(r)$ is the survey volume physically associated with the absorber as the function of impact parameters. We note that, to compare observations and simulations fairly, we obtained $n_0$ by integrating the observed [O III] luminosity from our detection limit to the bright end. Thus, we obtain $n_0 = n(L_{\rm [O\,III]} > 10^{42.3}\,{\rm erg\,s^{-1}}) \approx 1.873 \times 10^{-3}\,{\rm cMpc^{-3}}$. Here, we define the survey volume as different cylinders with radii ranging from 20 to 300 pkpc and heights of $|\Delta v| < 1000\,{\rm km\,s^{-1}}$. In Figure 4, we show the observed number abundance of galaxies surrounding this Mg II absorber. The errors are estimated by assuming Poissonian noise (84% confidence level) on the number of observed galaxies (Gehrels 1986).

In order to understand the relationship between metal absorbers and galaxies, as well as the environments in which they exist, we compare the observed correlation function with the predictions obtained from cosmological simulations. We focus on the Technicolor Dawn (Finlator et al. 2020), a hydrodynamic cosmological simulation for generating metal absorption and galaxies from $z = 5$ to 7 and taking into account radiative transfer effects. In Technicolor Dawn simulations, the correlation function is measured by linking the simulated Mg II absorber with similar EWs to our target (REW$_{2796} > 0.74$ Å) and these simulated galaxies with $M_{\rm UV} < -18$ (comparable to our [O III] detection limit). We show the comparison in Figure 4. The measured clustering effect between Mg II absorber and bright [O III] emitters is consistent with the simulated predictions. Our pilot study highlights that JWST NIRCam/WFSS observations offer an efficient method to understand the metal-enrichment process in the early Universe by detecting galaxies with $M_* \sim 10^9\,M_\odot$ and linking them with metal absorbers.

### 4.3. Constraints on Outflow Velocities

The metal-enrichment history of CGM is related to galactic outflows (Simcoe et al. 2004; Oppenheimer & Davé 2006).

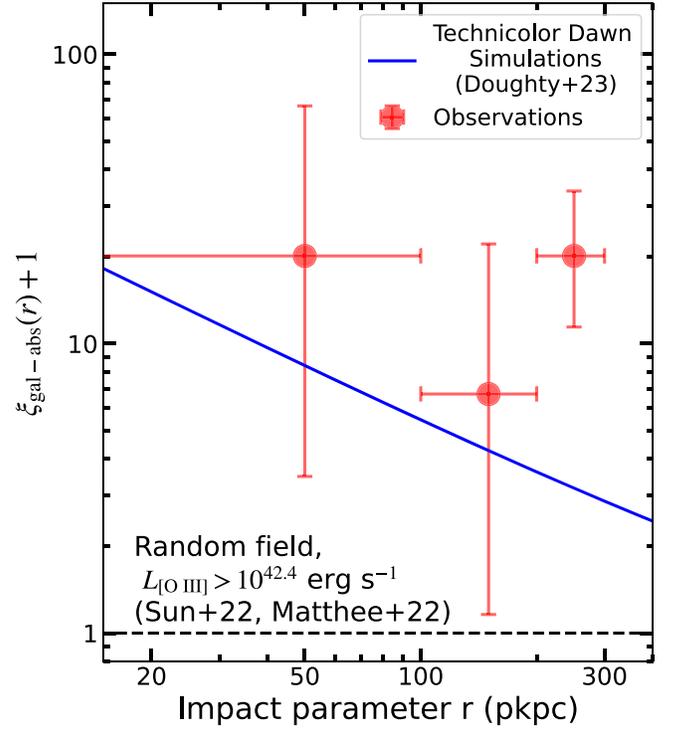

**Figure 4.** Number excess of galaxies around Mg II absorbers at $z \simeq 5$, $\xi_{\rm gal-abs}(r) = \frac{1}{n_0}\frac{N}{\Delta V} - 1$. The red dots are the measurements obtained from our observations. Error bars are estimated by assuming Poissonian uncertainties in the 84% confidence level (Gehrels 1986). The blue solid line indicates simulation-predicted values obtained from Doughty & Finlator (2023). The observed values are consistent with that predicted from cosmological simulations.

One key parameter that describes this process is the "outflow velocity" ($v_{\rm out}$). Nelson et al. (2019) calculated $v_{\rm out}$ from the TNG50 cosmological simulation. They found that, at $z \sim 5$, a galaxy with a stellar mass of $\sim 10^9\,M_\odot$ will have a typical outflow velocity of $v_{\rm out} \sim 250\,{\rm km\,s^{-1}}$ at an impact parameter of $\sim 20$ pkpc. Observationally, Díaz et al. (2021) estimated the mean outflow velocity to be $\approx 200\,{\rm km\,s^{-1}}$ based on the measured impact parameter between LAEs and C IV absorbers. In their measurement, they assumed that the metal-enriched gas starts being ejected at $z = 10$ since they do not know the star formation history of the galaxy (Figure 19 in their work). Similarly, Galbiati et al. (2023) also assumed a gas launching time and then estimated a mean outflow velocity of $\approx 200\,{\rm km\,s^{-1}}$.

From the SED fitting, we derived a stellar age of ($79^{+79}_{-57}$ Myr) by assuming a constant star-forming history for O3-038. Together with the observed impact parameter (24.9 pkpc), we estimate an outflow velocity of $v_{\rm out} = \frac{\rm impact\ parameter}{\rm stellar\ age} \approx 300\,{\rm km\,s^{-1}}$. This value is consistent with the simulated prediction (Muratov et al. 2015; Nelson et al. 2019).

### 5. Summary

In this Letter, we report the discovery of the host galaxy of a Mg II absorber at $z = 5.428$ using NIRCam/WFSS data from the JWST ASPIRE program (Wang et al. 2023). This galaxy is bright in [O III] with an [O III] luminosity of $L_{\rm [O\,III]} = 3.30 \pm 0.51 \times 10^{42}\,{\rm erg\,s^{-1}}$ and an [O III] REW of $490 \pm 80$ Å. Based on SED fitting to its HST+JWST photometry, we estimate the stellar mass, SFR, and stellar age of this absorber-selected galaxy as $M_* \simeq 10^{8.8^{+0.2}_{-0.4}} M_\odot$, SFR $\simeq 9.7^{+4.3}_{-2.7}\ M_\odot$ yr$^{-1}$, and Age $= 79^{+79}_{-57}$ Myr, respectively. The derived stellar mass and measured [O III] EW suggest that ASPIRE-J0305M31-O3-038 shares the same properties as other typical [O III] emitters at $z \sim 6$. Meanwhile, the SFR of this Mg II-selected galaxy is higher than the majority of metal-absorber-selected galaxies at $z \sim 5$ (Cai et al. 2017; Díaz et al. 2021; Wu et al. 2021). Furthermore, we find that the galaxy resides in a galaxy overdensity at $z \simeq 5.4$ with six additional galaxies located at $\Delta_v < 1000$ km s$^{-1}$. We measure the number excess of galaxies around the Mg II absorber in the radius range of 20–300 pkpc and find that it is consistent with that obtained from cosmological simulations (Doughty & Finlator 2023). This pilot experiment demonstrates the capability of NIRCam/WFSS in detecting the host galaxies of high-redshift metal absorbers. With the full data set of 25 quasar sight lines in the ASPIRE program, we expect to finally build up a statistical sample of absorber–galaxy pairs at $z > 5$ and constrain the metal enrichment in the early Universe.

## Acknowledgments

We thank Fuyan Bian, Mengtao Tang, and Jakob Helton for very helpful discussions. We thank Jorryt Matthee and the EIGER team for sharing their data and acknowledge Marta Galbiati and Michele Fumagalli for sharing their MUSE catalog.

This work is based on observations made with the NASA/ESA/CSA James Webb Space Telescope. The data were obtained from the Mikulski Archive for Space Telescopes at the Space Telescope Science Institute, which is operated by the Association of Universities for Research in Astronomy, Inc., under NASA contract NAS 5-03127 for JWST. These observations are associated with program #2078. Support for program #2078 was provided by NASA through a grant from the Space Telescope Science Institute, which is operated by the Association of Universities for Research in Astronomy, Inc., under NASA contract NAS 5-03127. Z.C., Y.W., X.L., Z.L., M.L., and S.Z. are supported by the National Key R&D Program of China (grant no. 2018YFA0404503), the National Science Foundation of China (grant no. 12073014). The science research grants from the China Manned Space Project with No. CMS-CSST2021-A05, and Tsinghua University Initiative Scientific Research Program (No. 20223080023) K.F. gratefully acknowledges support from STScI Program #HST-AR-16125.001-A. This program was provided by NASA through a grant from the Space Telescope Science Institute, which is operated by the Associations of Universities for Research in Astronomy, Incorporated, under NASA contract NAS5-26555. K.F.'s simulation utilized resources from the New Mexico State University High Performance Computing Group, which is directly supported by the National Science Foundation (OAC-2019000), the Student Technology Advisory Committee, and New Mexico State University and benefits from inclusion in various grants (DoD ARO-W911NF1810454; NSF EPSCoR OIA-1757207; Partnership for the Advancement of Cancer Research, supported in part by NCI grants U54 CA132383 (NMSU)). F.S. acknowledges support from the NRAO Student Observing Support (SOS) award SOSPA7-022. H.D.J. was supported by the National Research Foundation of Korea (NRF) funded by the Ministry of Science and ICT (MSIT) of Korea (No. 2020R1A2C3011091, 2021M3F7A1084525, 2022R1C1C2013543) G.B. was supported by the U.S. National Science Foundation through grant AST-1751404. SEIB acknowledges funding from the European Research Council (ERC) under the European Union's Horizon 2020 research and innovation program (grant agreement no. 740246 "Cosmic Gas"). E.P.F. is supported by the international Gemini Observatory, a program of NSF's NOIRLab, which is managed by the Association of Universities for Research in Astronomy (AURA) under a cooperative agreement with the National Science Foundation, on behalf of the Gemini partnership of Argentina, Brazil, Canada, Chile, the Republic of Korea, and the United States of America.

JWST data used in this Letter were obtained from the Mikulski Archive for Space Telescopes (MAST) at the Space Telescope Science Institute. The specific observations analyzed can be accessed via doi:10.17909/vt74-kd84.

*Facilities:* JWST, VLT:Kueyen, HST.

*Software:* astropy (Astropy Collaboration et al. 2013, 2018, 2022), PypeIt (Prochaska et al. 2020), VoigtFit (Krogager 2018), BEAGLE (Chevallard & Charlot 2016), drizzlepac (STSCI Development Team 2012).

## Appendix
## SED-fitting Result

To obtain the stellar properties, we fit the HST + JWST photometries with Bayesian Analysis of Galaxy SEDS (BEAGLE; Chevallard & Charlot 2016). We assume a constant star formation history and apply a flat prior in the ranges of metallicity ($0.01 < Z/Z_\odot < 0.1$) and stellar mass ($5 \leqslant \log(M_*/M_\odot) \leqslant 12$). For more details, we refer interested readers to Section 3.1 in Lin et al. (2023). The best-fit SED fitting results are shown in Figure 5 and listed in Table 1. We note that the absolute UV magnitude ($M_{\rm UV}$) is derived from the posterior SED by integrating the model spectra with a 100 Å width window at rest frame 1500 Å. Following Tacchella et al. (2023), we also include a noise floor of 5% in our SED-fitting results.

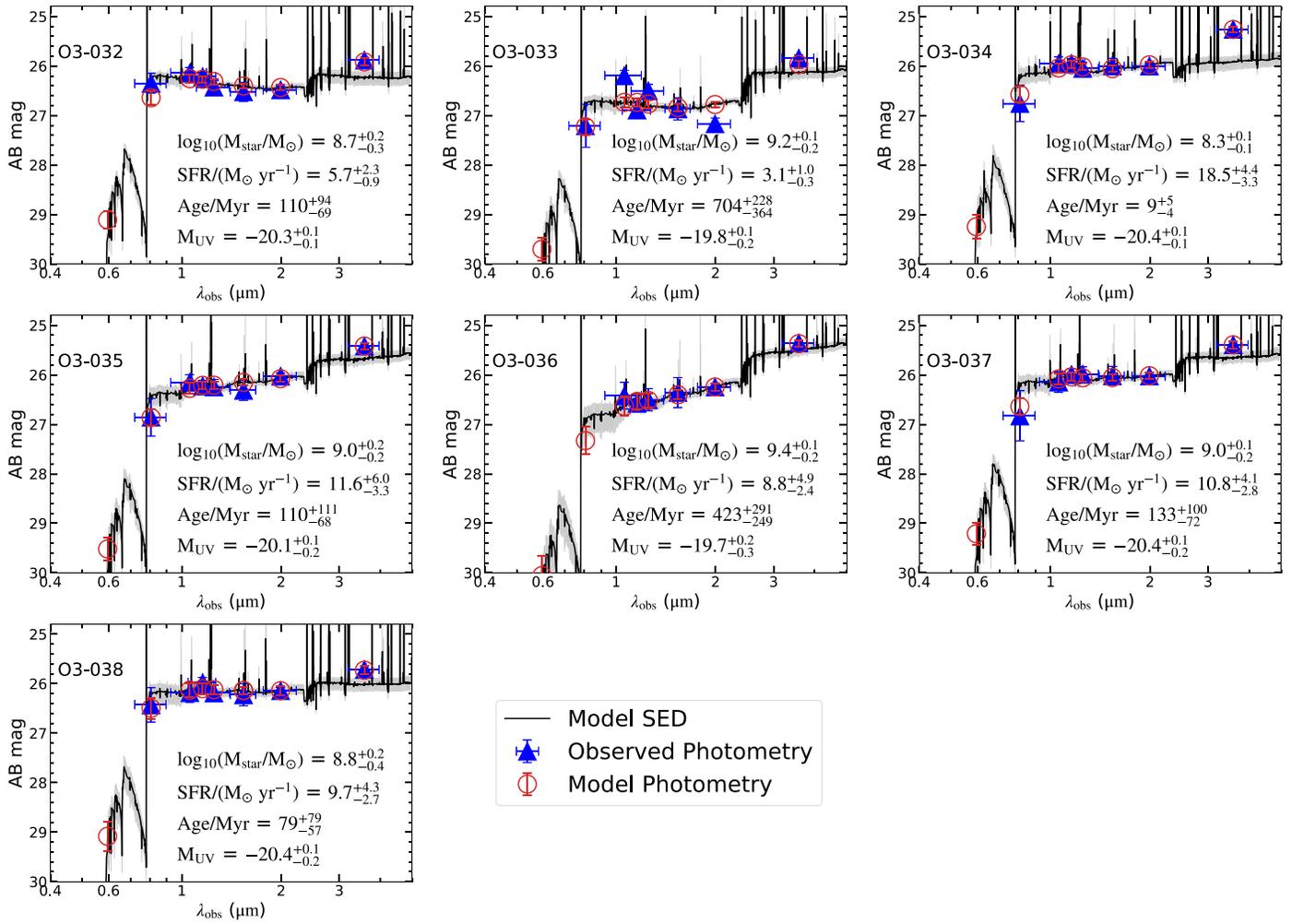

**Figure 5.** Best SED solution based on BEAGLE. The photometric data points from HST+ JWST are represented by blue triangles, while the corresponding model photometries are shown in red. The median value of the posterior distribution of SEDs is shown with the black line. The gray region represents the 16th and 84th confidence intervals.

### ORCID iDs

Yunjing Wu ● https://orcid.org/0000-0003-0111-8249
Feige Wang ● https://orcid.org/0000-0002-7633-431X
Zheng Cai ● https://orcid.org/0000-0001-8467-6478
Xiaohui Fan ● https://orcid.org/0000-0003-3310-0131
Kristian Finlator ● https://orcid.org/0000-0002-0496-1656
Jinyi Yang ● https://orcid.org/0000-0001-5287-4242
Joseph F. Hennawi ● https://orcid.org/0000-0002-7054-4332
Fengwu Sun ● https://orcid.org/0000-0002-4622-6617
Jaclyn B. Champagne ● https://orcid.org/0000-0002-6184-9097
Xiaojing Lin ● https://orcid.org/0000-0001-6052-4234
Zihao Li ● https://orcid.org/0000-0001-5951-459X
Zuyi Chen ● https://orcid.org/0000-0002-2178-5471
Eduardo Bañados ● https://orcid.org/0000-0002-2931-7824
George D. Becker ● https://orcid.org/0000-0003-2344-263X
Sarah E. I. Bosman ● https://orcid.org/0000-0001-8582-7012
Gustavo Bruzual ● https://orcid.org/0000-0002-6971-5755
Stephane Charlot ● https://orcid.org/0000-0003-3458-2275
Hsiao-Wen Chen ● https://orcid.org/0000-0001-8813-4182
Jacopo Chevallard ● https://orcid.org/0000-0002-7636-0534
Anna-Christina Eilers ● https://orcid.org/0000-0003-2895-6218
Emanuele Paolo Farina ● https://orcid.org/0000-0002-6822-2254
Xiangyu Jin ● https://orcid.org/0000-0002-5768-738X
Hyunsung D. Jun ● https://orcid.org/0000-0003-1470-5901
Koki Kakiichi ● https://orcid.org/0000-0001-6874-1321
Mingyu Li ● https://orcid.org/0000-0001-6251-649X
Weizhe Liu ● https://orcid.org/0000-0003-3762-7344
Maria A. Pudoka ● https://orcid.org/0000-0003-4924-5941
Wei Leong Tee ● https://orcid.org/0000-0003-0747-1780
Zhang-Liang Xie ● https://orcid.org/0000-0002-0125-6679
Siwei Zou ● https://orcid.org/0000-0002-3983-6484

### References

Astropy Collaboration, Price-Whelan, A. M., Lim, P. L., et al. 2022, ApJ, 935, 167
Astropy Collaboration, Price-Whelan, A. M., Sipőcz, B. M., et al. 2018, AJ, 156, 123
Astropy Collaboration, Robitaille, T. P., Tollerud, E. J., et al. 2013, A&A, 558, A33
Becker, G. D., Pettini, M., Rafelski, M., et al. 2019, ApJ, 883, 163
Bertin, E., Schefer, M., Apostolakos, N., et al. 2020, in ASP Conf. Ser. 527, Astronomical Data Analysis Software and Systems XXIX, ed. R. Pizzo et al. (San Francisco, CA: ASP), 461
Bordoloi, R., Simcoe, R. A., Matthee, J., et al. 2023, arXiv:2307.01273

Brinchmann, J., Charlot, S., White, S. D. M., et al. 2004, MNRAS, 351, 1151
Bushouse, H., Eisenhamer, J., Dencheva, N., et al. 2022, JWST Calibration Pipeline, v1.8.2, Zenodo, doi:10.5281/zenodo.7325378
Cai, Z., Fan, X., Dave, R., Finlator, K., & Oppenheimer, B. 2017, ApJL, 849, L18
Champagne, J. B., Casey, C. M., Finkelstein, S. L., et al. 2023, ApJ, 952, 99
Chen, S.-F. S., Simcoe, R. A., Torrey, P., et al. 2017, ApJ, 850, 188
Chen, Z., Stark, D. P., Endsley, R., et al. 2023, MNRAS, 518, 5607
Chevallard, J., & Charlot, S. 2016, MNRAS, 462, 1415
Cooper, T. J., Simcoe, R. A., Cooksey, K. L., et al. 2019, ApJ, 882, 77
Davies, R. L., Ryan-Weber, E., D'Odorico, V., et al. 2023, MNRAS, 521, 289
Díaz, C. G., Koyama, Y., Ryan-Weber, E. V., et al. 2014, MNRAS, 442, 946
Díaz, C. G., Ryan-Weber, E. V., Cooke, J., Koyama, Y., & Ouchi, M. 2015, MNRAS, 448, 1240
Díaz, C. G., Ryan-Weber, E. V., Karman, W., et al. 2021, MNRAS, 502, 2645
D'Odorico, V., Finlator, K., Cristiani, S., et al. 2022, MNRAS, 512, 2389
Doughty, C. C., & Finlator, K. M. 2023, MNRAS, 518, 4159
Endsley, R., Stark, D. P., Chevallard, J., & Charlot, S. 2021, MNRAS, 500, 5229
Evrard, A. E., Bialek, J., Busha, M., et al. 2008, ApJ, 672, 122
Farina, E. P., Venemans, B. P., Decarli, R., et al. 2017, ApJ, 848, 78
Ferragamo, A., Rubiño-Martín, J. A., Betancort-Rijo, J., et al. 2020, A&A, 641, A41
Finlator, K., Doughty, C., Cai, Z., & Díaz, G. 2020, MNRAS, 493, 3223
Finlator, K., Muñoz, J. A., Oppenheimer, B. D., et al. 2013, MNRAS, 436, 1818
Galbiati, M., Fumagalli, M., Fossati, M., et al. 2023, MNRAS, 524, 3473
Gehrels, N. 1986, ApJ, 303, 336
Greif, T. H., Glover, S. C. O., Bromm, V., & Klessen, R. S. 2010, ApJ, 716, 510
Hirschmann, M., Naab, T., Davé, R., et al. 2013, MNRAS, 436, 2929
Horne, K. 1986, PASP, 98, 609
Kashino, D., Lilly, S. J., Matthee, J., et al. 2023a, ApJ, 950, 66
Kashino, D., Lilly, S. J., Simcoe, R. A., et al. 2023b, Natur, 617, 261
Kauffmann, G., Heckman, T. M., White, S. D. M., et al. 2003, MNRAS, 341, 33
Keating, L. C., Puchwein, E., Haehnelt, M. G., Bird, S., & Bolton, J. S. 2016, MNRAS, 461, 606
Krogager, J.-K. 2018, arXiv:1803.01187
Kümmel, M., Bertin, E., Schefer, M., et al. 2020, in ASP Conf. Ser. 527, Astronomical Data Analysis Software and Systems XXIX, ed. R. Pizzo (San Francisco, CA: ASP), 29
Lin, X., Cai, Z., Zou, S., et al. 2023, ApJL, 944, L59
Ma, X., Hopkins, P. F., Garrison-Kimmel, S., et al. 2018, MNRAS, 478, 1694
Mannucci, F., Cresci, G., Maiolino, R., Marconi, A., & Gnerucci, A. 2010, MNRAS, 408, 2115
Matejek, M. S., & Simcoe, R. A. 2012, ApJ, 761, 112
Matthee, J., Mackenzie, R., Simcoe, R. A., et al. 2023, ApJ, 950, 67
Meyer, R. A., Bosman, S. E. I., Kakiichi, K., & Ellis, R. S. 2019, MNRAS, 483, 19
Munari, E., Biviano, A., Borgani, S., Murante, G., & Fabjan, D. 2013, MNRAS, 430, 2638
Muratov, A. L., Kereš, D., Faucher-Giguère, C.-A., et al. 2015, MNRAS, 454, 2691
Nelson, D., Pillepich, A., Springel, V., et al. 2019, MNRAS, 490, 3234
Oesch, P. A., Brammer, G., Naidu, R. P., et al. 2023, MNRAS, 525, 2864
Oppenheimer, B. D., & Davé, R. 2006, MNRAS, 373, 1265
Oppenheimer, B. D., Davé, R., & Finlator, K. 2009, MNRAS, 396, 729
Ouchi, M., Ono, Y., & Shibuya, T. 2020, ARA&A, 58, 617
Prochaska, J., Hennawi, J., Westfall, K., et al. 2020, JOSS, 5, 2308
Rieke, M. J., Kelly, D. M., Misselt, K., et al. 2023, PASP, 135, 028001
Savage, B. D., & Sembach, K. R. 1991, ApJ, 379, 245
Schaerer, D., Ginolfi, M., Béthermin, M., et al. 2020, A&A, 643, A3
Schindler, J.-T., Farina, E. P., Bañados, E., et al. 2020, ApJ, 905, 51
Simcoe, R. A., Sargent, W. L. W., & Rauch, M. 2004, ApJ, 606, 92
Sorini, D., Davé, R., & Anglés-Alcázar, D. 2020, MNRAS, 499, 2760
Steidel, C. C., Erb, D. K., Shapley, A. E., et al. 2010, ApJ, 717, 289
STSCI Development Team, 2012 DrizzlePac: HST image software, Astrophysics Source Code Library, ascl:1212.011
Sun, F., Egami, E., Pirzkal, N., et al. 2022, ApJL, 936, L8
Sun, F., Egami, E., Pirzkal, N., et al. 2023, ApJ, 952, 53
Sun, G., & Furlanetto, S. R. 2016, MNRAS, 460, 417
Tacchella, S., Eisenstein, D. J., Hainline, K., et al. 2023, ApJ, 952, 74
Tang, M., Stark, D. P., Chevallard, J., & Charlot, S. 2019, MNRAS, 489, 2572
Tumlinson, J., Peeples, M. S., & Werk, J. K. 2017, ARA&A, 55, 389
Venemans, B. P., Findlay, J. R., Sutherland, W. J., et al. 2013, ApJ, 779, 24
Wang, F., Yang, J., Hennawi, J. F., et al. 2023, ApJL, 951, L4
Whitler, L., Stark, D. P., Endsley, R., et al. 2023, MNRAS, 519, 5859
Wise, J. H., Turk, M. J., Norman, M. L., & Abel, T. 2012, ApJ, 745, 50
Wu, Y., Cai, Z., Neeleman, M., et al. 2021, NatAs, 5, 1110
Yang, J., Wang, F., Fan, X., et al. 2023, ApJL, 951, L5
Zou, S., Jiang, L., Shen, Y., et al. 2021, ApJ, 906, 32